\documentstyle[12pt]{article}
\begin{document}
\baselineskip 15pt

\begin{titlepage}
\rightline{MC/TH 96/14}
\rightline{TPR-96-06}
\rightline{May 1996}

\vskip 1.5cm
\centerline{\large\bf 
Comment on ``Relativistic kinetic equations for electromagnetic,}
\vskip 0.3cm
\centerline{\large\bf scalar and pseudoscalar interactions'' }
\vskip 1.0cm
\centerline{Abdellatif Abada and   Michael C. Birse}
\vskip 0.3cm
\centerline{\it Theoretical Physics Group, Department of Physics and 
Astronomy,}
\centerline{\it University of Manchester, Manchester M13 9PL, UK}

\vskip 1cm
\centerline{Pengfei Zhuang}
\vskip 0.3cm
\centerline{\it Gesellschaft f\"ur Schwerionenforschung, Theory Group, }
\centerline{\it P.O.Box 110552, D-64220 Darmstadt, Germany}
\vskip 1cm
\centerline{Ulrich Heinz$^*$}
\centerline{\it Department of Physics, Duke University, Durham, N.C. 
27708-0305, USA}

\vskip 2.0cm
\centerline{\large \bf Abstract}
\vskip 0.5cm
\noindent

It is found that the extra quantum constraints to the spinor components of 
the equal-time Wigner function given in a recent paper by Zhuang and Heinz 
should vanish identically. We point out here the origin of the error and give
an interpretation of the result. However, the principal idea of obtaining a
complete equal-time transport theory by energy averaging the covariant theory
remains valid. The classical transport equation for the spin density is also
found to be incorrect. We give here the correct form of that equation and
discuss briefly its structure.

\end{titlepage}
\baselineskip 0.8 cm

In recent papers \cite{ZH1,ZH2} two of us (P.Z. and U.H.) investigated the
equal-time transport theory for a system with electromagnetic, scalar and 
pseudoscalar interactions by taking the energy average of the corresponding
covariant theory. It was shown that the spinor components of the equal-time
Wigner function, which are the zeroth-order energy moments of the 
corresponding components of the covariant Wigner function, are coupled to 
the first-order moments and satisfy the generalized BGR equations \cite{BGR}.
When confirming these conclusions by an independent check of the calculations 
in \cite{ZH2} we found, however, that the extra quantum constraints (ZH21) on
the equal-time components should vanish identically. Furthermore we found a
related error in the classical transport equation (ZH18) for the spin density.
(We refer to specific equations from Ref.~\cite{ZH2} by adding ZH in front of
the equation number.) We here point out the origin of the error and give the
correct derivation.

In \cite{ZH2} Eqs.~(ZH21) were derived by eliminating the first-order 
energy moments from the constraint equations (ZH10) by combining them with 
the BGR transport equations (ZH9) and with the first-order energy moments
of those covariant equations (ZH7) whose zeroth-order moments gave rise to the
BGR equations. The mathematical mistake which leads to Eqs.~(ZH21) is that
the second term on the r.h.s. of
 \begin{equation}
 \label{1}
    \partial_p^n \left(p W\right) = 
    p\partial_p^n W + n\partial_p^{n-1} W
\end{equation}
was inadvertently dropped for $n\geq 2$. In this formula $W$ stands for 
the equal-time Wigner function or any of its spinor components, and 
$\partial_p^n$ is the $n$th-order momentum derivative which appears
in the electromagnetic, scalar and pseudoscalar field operators ${\bf E}, 
{\bf B}, \sigma_e, \sigma_o, \pi_e$ and $\pi_o$ defined in \cite{ZH2}. 
This meant that, for instance, the second term was omitted from
\begin{equation}
\sigma_e {\bf p}W={\bf p}\sigma_e W -\hbar 
(\mbox{\boldmath $\nabla$} \sigma_o) W/2,
\end{equation}
and related formulae. The effects of these terms in the field operators are
to cancel exactly the terms in Eqs.~(ZH21) involving the operators ${\bf M},
{\bf L}, {\bf F}_{\sigma_o}, {\bf F}_{\sigma_e}$ and ${\bf F}_\sigma, {\bf
F}_{\pi_o}$ and ${\bf F}_{\pi_e}$ and ${\bf F}_\pi$. Therefore no extra
constraints on the equal-time Wigner function arise from equations (ZH10).

It is possible to give a deeper interpretation of this result: In \cite{ZH2} 
the two groups of fundamental equal-time kinetic equations are the BGR 
transport equations (ZH9) and the constraint equations (ZH10). Eqs.~(ZH9) 
determine the evolution of the zeroth-order moments, while Eqs.~(ZH10) give 
explicit expressions for the first-order moments in terms of the zeroth-order
ones. In principle, another group of equations which connects zeroth- and
first-order energy moments can be derived from the first-order energy moment
of the covariant version of the BGR equations. The above calculation shows
that this additional set of equations contains no independent information;
Eqs.~(ZH9) and (ZH10) are the only independent equations controlling the
behavior of the zeroth- and first-order energy moments.

Furthermore, quite generally it is impossible to extract any extra
relationships among the zeroth-order moments from the constraints (ZH10),
except in the classical limit. In this limit, the covariant components satisfy
the mass-shell constraints $p^2=E^2-{\bf p}^2=(m^*)^2$, and their energy
dependence degenerates to two delta-functions at $E=\pm E_p=\sqrt{{\bf
p}^2+(m^*)^2}$, with $m^*$ being the constituent quark mass (see Eq.\ 
(\ref{mstar}) below). In this case (and only in this case) all higher energy
moments are algebraically related to the zeroth order moment, and in particular
 \begin{equation}
\label{2}
W_1^\pm(x,{\bf p}) = \int d p_0 p_0 W^\pm(x,p) = \pm E_p W^\pm(x,{\bf p})\ .
 \end{equation}
This extra relationship between the classical limits of the first and 
zeroth order energy moments turns the constraint equations (ZH10) into
a set of essential constraints on the classical transport equations (ZH9),
which allow one to reduce the number of independent distribution functions by 
a factor of two. However, this works only in the classical limit, and no
such constraints can be derived in the general quantum case. 

When redoing the calculations a related error was also discovered in the
classical transport equation (ZH18) for the spin density ${\bf g}_0$
This should read
\begin{eqnarray}
&& \left[\partial_t+{{\bf p}\over E_p}{\cdot}\mbox{\boldmath $\nabla$}
+\left(e{\bf E}+e {{\bf p}\over E_p}\times{\bf B}
-\mbox{\boldmath $\nabla$}E_p\right){\cdot}\mbox{\boldmath $\partial$}_p
\right]{\bf g_0} \label{g0}\\
&&\qquad ={e\over E_p^2}{\bf p}\times({\bf E}\times{\bf g_0})
-{e\over E_p}{\bf B}\times {\bf g_0} 
-{1\over (m^*)^2}\left({\dot E_p\over E_p}{\bf p}+
\mbox{\boldmath $\nabla$}
E_p \right)\times ({\bf p}\times {\bf g_0}) \nonumber\\
&& \qquad\qquad+{1\over E_p(m^*)^2}\left({\cal A}_0~{\bf p}\times {\bf g_0}+
{\bf {\cal A}}\times\left(E_p{\bf g_0}
-{{\bf p}\over E_p}({\bf p}{\cdot}{\bf g_0})
\right)\right)\ ,\nonumber
\end{eqnarray} 
where $m^*$ is defined by 
\begin{equation}
m^{*2}=(m-V_\sigma)^2+V_\pi^2,
\label{mstar}
\end{equation}
and ${\cal A}_{\mu}$ has a form similar to the mesonic axial current of the
linear sigma model\cite{GML}
\begin{equation}
{\cal A}_{\mu}= (m-V_{\sigma})\partial_{\mu}V_{\pi} 
- V_{\pi} \partial_{\mu} (m-V_{\sigma}) .
\end{equation}
Here the scalar and pseudoscalar potentials are related to the corresponding
fields by\cite{ZH2}
\begin{equation}
V_\sigma=g\sigma, \qquad V_\pi=g\pi.
\end{equation}

The equation (\ref{g0}) has a rather natural structure involving the mass 
$m^*$, which is chirally invariant if the current mass $m$ vanishes, and the
axial current ${\cal A}_\mu$, which gives rise to the classical analogue of
the familiar derivative coupling of the pions. The corresponding covariant
equation has been derived by Florkowski {\it et al.}\cite{FHKN} for particles
in the presence of mean scalar and pion fields. It can be seen to contain
terms with similar structures to those in Eq.~(\ref{g0}). In particular the
term involving the gradient of the chiral angle gives rise to terms involving
${\cal A}^\mu$ when integrated over energy. These terms are essential to
maintaining PCAC for fermions whose masses are generated by spontaneous
breaking of chiral symmetry, as in the linear sigma\cite{GML} and
Nambu--Jona-Lasinio models\cite{NJL}.

Note that in the classical limit the equations for the density (ZH17) and spin
density (\ref{g0}) decouple. This is because the spin of the fermions is of
order $\hbar$. Hence, although spin-orbit interactions show up in the 
spin-precession terms of Eq.~(\ref{g0}), they have no effect on the spatial
motion of the particles at the classical level. Indeed at first order in
$\hbar$ the quantum corrections arise entirely from spin-orbit coupling and so
in the particular case of scalar QED there are no corrections at this
order \cite{ZH1}.

\vskip 1cm
\centerline {\bf ACKNOWLEDGMENTS}
U.H. is grateful to Prof.~B. M\"uller and the Duke University Physics
Department for their warm hospitality. His work was supported financially by
GSI, DFG, and BMBF. A.A. and M.C.B. acknowledge support from the EPSRC and
PPARC.

\end{document}